\documentclass[twocolumn,
	prd,aps,superscriptaddress,nofootinbib]{revtex4-1}

\usepackage{graphicx}  % needed for figures
\usepackage{xcolor}
\usepackage{dcolumn}   % needed for some tables
\usepackage{bm}        % for math
\usepackage{amssymb}   % for math
\usepackage{amsmath}
\usepackage{verbatim}
\usepackage{dcolumn}
\usepackage[
pdfauthor={Nirmal Raj}]{hyperref}

\definecolor{nirmal}{rgb}{1,0,1}

\newcommand{\NS}{neutron star }

\renewcommand{\eqref}[1]{\ref{#1}} % eqref vs ref

\begin{document}

\preprint{UCR-TR-2019-FLIP-NCC-1701-B}

\title{Relativistic capture of dark matter by electrons in neutron stars}
\author{Aniket Joglekar}
\email{aniket@ucr.edu}
\affiliation{Department of Physics and Astronomy, University of California, Riverside, California 92521, USA}
\author{Nirmal Raj}
\email{nraj@triumf.ca} 
\affiliation{TRIUMF, 4004 Wesbrook Mall, Vancouver, BC V6T 2A3, Canada}
\author{Philip Tanedo}
\email{flip.tanedo@ucr.edu} 
\affiliation{Department of Physics and Astronomy, University of California, Riverside, California 92521, USA}
\author{Hai-Bo Yu}
\email{haiboyu@ucr.edu} 
\affiliation{Department of Physics and Astronomy, University of California, Riverside, California 92521, USA}

\date{\today}

\begin{abstract}
	Dark matter can capture in neutron stars and heat them to observable luminosities. We study relativistic scattering of dark matter on highly degenerate electrons. We develop a Lorentz invariant formalism to calculate the capture probability of dark matter that accounts for the relativistic motion of the target particles and Pauli exclusion principle. We find that the actual capture probability can be five orders of magnitude larger than the one estimated using a nonrelativistic approach. For dark matter masses $10~{\rm eV}\textup{--}10~{\rm PeV}$, neutron star heating complements and can be more sensitive than terrestrial direct detection searches. The projected sensitivity regions exhibit characteristic features that demonstrate a rich interplay between kinematics and Pauli blocking of the DM--electron system. Our results show that old neutron stars could be the most promising target for discovering leptophilic dark matter.

\end{abstract}

\maketitle
Dark matter (DM) makes up more than $80\%$ of the mass in the universe, but its identity remains largely unknown. There has been growing interest in signals of DM capture in compact stars~\cite{Goldman:1989nd,Gould:1989gw,Kouvaris:2007ay,Bertone:2007ae,deLavallaz:2010wp,Kouvaris:2010vv,McDermott:2011jp,Kouvaris:2011fi,Guver:2012ba,Bramante:2013hn,Bell:2013xk,Bramante:2013nma,Kouvaris:2010jy,McCullough:2010ai,Bertoni:2013bsa,Perez-Garcia:2014dra,Bramante:2015cua,Graham:2015apa,Cermeno:2016olb,Krall:2017xij,Graham:2018efk,McKeen:2018xwc,Acevedo:2019gre,Janish:2019nkk,Bell:2020jou}. In particular, neutron stars have super-nuclear densities that make them intriguing DM detectors. Incident DM particles are accelerated by the steep gravitational potential and may deposit their kinetic energy as heat via scattering with individual stellar constituents~\cite{Baryakhtar:2017dbj,Raj:2017wrv,Bell:2018pkk,Garani:2018kkd,Chen:2018ohx,Hamaguchi:2019oev,Camargo:2019wou,Bell:2019pyc,Garani:2019fpa,Acevedo:2019agu}\footnote{If DM were made of primordial black holes, they could be slowed down and captured in the stellar medium via the effect of dynamical friction, see e.g.~\cite{Montero_Camacho_2019}}. If radio telescopes observe a nearby old pulsar, upcoming infrared telescopes may measure the stellar luminosity and {detect} this DM kinetic heating. This search is largely independent of the details of DM interactions with Standard Model particles and thus sensitive to numerous scenarios of DM that are otherwise inaccessible to terrestrial detectors~\cite{Baryakhtar:2017dbj,Raj:2017wrv,Bell:2018pkk,Acevedo:2019agu}. 

The electron--DM portal is a well-motivated scenario that is crucial for light DM detection; see~\cite{Battaglieri:2017aum}. There have been a wide-ranging suite of experimental efforts in this new direction~\cite{Angle:2011th,Agnes:2018oej,Agnese:2018col,Abramoff:2019dfb,Aprile:2019xxb,Aguilar-Arevalo:2019wdi,Aprile:2019jmx,Arnaud:2020svb,x1texcess}. In this \textit{Letter}, we show that despite making up only $\sim3\times10^{-3}\,\%$ of the stellar mass, the electrons in a neutron star are excellent targets for capturing DM. Neutron star heating can search for DM masses and couplings that greatly exceed the limits set by the Earth-based direct detection experiments. 

Electrons in the neutron star are ultrarelativistic, highly degenerate and are moving in random directions, while DM particles approaching a neutron star are quasirelativistic with star escape velocity $v_\text{esc} \sim 0.6$.  Because each DM--electron center of momentum frame is distinct and highly boosted from the neutron star frame, the conventional formalism, developed for nonrelativistic targets, is invalid in calculating the capture probability. For the system we consider, it is necessary to specify the key scattering ingredients in different reference frames. The DM--electron scattering cross section is most conveniently expressed in the center of momentum frame of each DM--electron pair, while the target Fermi--Dirac distributions are best defined in the \NS frame.

We develop a manifestly Lorentz invariant formalism to express the capture probability per DM particle in the \NS in terms of the kinematic ingredients discussed above. It incorporates Pauli blocking and other capture conditions so that one may integrate over the phase space available for DM capture. We apply this formalism to {two benchmark DM scenarios} and estimate sensitivities on model parameters from \NS heating. The first assumes a contact operator to model DM--electron interactions. The second contains a light mediator particle with fixed in-medium effective masses of $1~{\rm keV}$ and $10~{\rm MeV}$, well below the Fermi momentum.

We find that the actual electron capture probability can be a factor of $(p_{\rm F}/m_e)^2\sim10^5$ larger than the estimate using a nonrelativistic approach. For DM masses between $10~{\rm eV}\textup{--}10~{\rm PeV}$, the \NS constraints are stronger than current limits from DM direct detection experiments in most of the mass range including the light DM regime. In particular, neutron star heating could be the most promising method to discover leptophilic DM. 

\vspace{.5em}
{\noindent\bf Lorentz-invariant capture.} 
A DM particle is bound to a \NS if it loses its halo kinetic energy $E_{\rm halo}=m_\chi v^2_{\rm h}/2$ by scattering within the star. For $N_\text{hit}$ scatters that deposit average energy $\left<\Delta E\right>$, capture occurs when $N_{\rm hit} \left<\Delta E\right> > E_{\rm halo}$. We take the DM velocity in the halo to be $v_{\rm h}=220~{\rm km/s}$. The rate of kinetic energy deposition is $\dot K =\left(\gamma_{\rm esc}-1\right)\dot{m}_\chi\,f$, where $\gamma_{\rm esc} = (1-v^2_{\rm esc})^{-1/2}$, $\dot{m}_\chi \sim 10^{25}$~GeV/s is the mass capture rate, and $f$ is the optical depth of DM in the star such that the probability for a transiting DM particle to capture is given by $1-e^{-f}$; as we will be concerned with the optically thin limit, we treat $f$ as the capture probability. 
This process equilibrates on galactic timescales and the deposited energy is radiated as heat.  The resulting blackbody temperature is $T_\star\approx{1600}~f^{1/4}~\rm{K}$~\cite{Baryakhtar:2017dbj,Raj:2017wrv}. For $f=1$ this is $\mathcal{O}(10)$ higher than that of a $10^9$ year-old \NS that is not heated by DM~\cite{Page:2004fy,Yakovlev:2004iq}, unless the \NS undergoes rotochemical heating that depends on the initial period and nuclear modelling~\cite{Hamaguchi:2019oev}. The key step to accurately study DM signals from neutron star heating is to calculate the capture probability per DM particle, $f$.

To develop a formalism for ${f}$ that is manifestly Lorentz invariant, we first consider the frame-invariant number of scattering events ($d\nu$) constructed in the DM rest frame in which the cross section and relative velocity can be properly defined~\cite{LL}:
\begin{align}
 d\nu=(d\sigma\cdot v\cdot dn_{\rm T}\cdot\Delta t\cdot dn_\chi\cdot\Delta V)_{\rm DM},
 \label{eq:eq1}
\end{align}
where
$d\sigma $ is the cross section, $v$ is the relative velocity,
$dn_\chi$, $dn_{\rm T}$ are infinitesimal DM and target number densities respectively, $\Delta V$ denotes interaction volume and $\Delta t$ {transit} time; all evaluated in the DM frame. 
Since $d\nu$ and $dn_\chi\Delta V$ are Lorentz invariant,  {so is} their ratio $d{f}=d\nu/(\Delta V dn_\chi)=(d\sigma\cdot v\cdot dn_{\rm T}\cdot\Delta t)_{\rm DM} $, the infinitesimal scattering probability. So we can write $f$ in terms of the corresponding variables in the \NS frame $d{f} =(d\sigma\cdot v\cdot dn_{\rm T}\cdot\Delta t)_{\rm NS}$. 
For a given target 4-momentum $p_\mu=(E_p,\vec{p})_{\rm NS}$ and DM 4-momentum $k_\mu=(E_k,\vec{k})_{\rm NS}$ in the \NS frame, there exists a relation, $(d\sigma\cdot v)_{\rm NS}=(d\sigma)_{\rm DM}(v_\text{M\o l})_{\rm NS}$, where $(v_\text{M\o l})_{\rm NS}=\sqrt{(p\cdot k)^2-m^2_{\rm T}m^2_\chi}/{(E_p E_k)_{\rm NS}}$ is the M\o ller velocity in the \NS frame. 
From this and using the fact that the cross section is invariant under boost along the collision axis, i.e., $(d\sigma)_{\rm DM}=(d\sigma)_{\rm CM}$, where ``CM" denotes the center of momentum frame, we obtain an expression for ${df}$,
\begin{align}
d{f} =\left(\frac{d\sigma}{d\Omega}\right)_{\rm CM}d\Omega_{\rm CM}(v_\text{M\o l} dn_{\rm T}\Delta t)_{\rm NS},
\label{eq:f}
\end{align}
where $d\Omega_{\rm CM}=d\cos\psi\, d\alpha$, for CM polar and azuimuthal angles $\psi$ and $\alpha$. Note that the last term in parentheses is Lorentz invariant. For what follows, we will suppress subscript ``NS" when referencing a variable in the \NS frame, except in a few instances to avoid confusion. 

\vspace{.5em}
{\noindent\bf Pauli blocking and phase space.} 
To evaluate ${f}$ in Eq.~\eqref{eq:f}, we need to perform the phase-space integral over $d\Omega_{\rm CM}\,dn_{\rm T}$. However, not all parts of the phase space are allowed to interact due to the Pauli exclusion principle, which requires the target particle to be knocked out of its Fermi sea in order to interact. Making use of the Lorentz invariance of ${f}$, we analyze the Pauli blocking condition in the \NS frame, where the Fermi surface is spherical. The condition can be expressed in the form of the Heaviside step function $\Theta(\Delta E +E_p - E_F )$, where $E_F$ is the Fermi energy and $\Delta E$ is the energy transferred to the target in the collision; both of them are in the \NS frame. 
Note that $\Delta E$ is related to the momentum transfer in the CM frame ($\vec{q}_{\rm CM}$) as $\Delta E=\vec{\beta}_{\rm CM}\cdot\vec{q}_{\rm CM}/\sqrt{1-\beta^2_{\rm CM}}$, where $\vec{\beta}_{\rm CM}=(\vec{p}+\vec{k})/(E_p+E_k)$ is the boost from the \NS to the CM frame. Finally, we must satisfy the capture condition, $N_{\rm hit}\left<\Delta E\right>>E_{\rm halo}$. This is done by summing over $N_{\rm hit}$ in a conservative way to ensure at least $E_{\rm halo}$ is transferred to the \NS during transit of a DM particle through it. {This accounts for the case when many scatters with smaller $\Delta E$ are more efficient than a single scatter with large $\Delta E$.} Putting these together, we have 
\small
\begin{widetext}
	\begin{align}
	 {f}= 	 \sum
	 \limits_{N_{\rm hit}\,\in\,\mathbb{Z}}
	  	\frac{\left<n_{\rm T}\right>\Delta t}{N_{\rm hit}}
	 	\int d\Omega_\text{NS}
	 	\int\limits^{p_\text{F}}_0 d|\bar{p}|\frac{|\bar{p}|^2}{V_\text{F}} v_\text{M\o l}
	 			\int d\Omega_\text{CM}\left(\frac{d\sigma}{d\Omega}\right)_\text{CM}\Theta\left(\Delta E+E_p-E_\text{F}\right)\Theta\left(\frac{E_{\rm halo}}{N_{\rm hit}-1}-\Delta E\right)\Theta\left(\Delta E - \frac{E_{\rm halo}}{N_{\rm hit}} \right) 
	 	\ ,
  \label{eqn:fullf}
\end{align}
\end{widetext}
\normalsize
where $\left<n_{\rm T}\right>$ is the average number density of the target species in the \NS core, $V_{\rm F}=4\pi p^3_{\rm F}/3$ is the Fermi volume, and $dn_{\rm T}=|\vec{p}|^2d|\vec{p}|\,d\Omega_{\rm NS}/V_{\rm F}$. 
We take $\left<n_{\rm T}\right>=3M_\star Y_{\rm T}/4\pi m_{\rm n} R^3_\star$, where $Y_{\rm T}$ is the target's volume-averaged number per nucleon, $M_\star$ is the mass of the \NS and $R_\star$ its radius.
For the constituents $\{e^-, \mu^-, p^+, n\}$, we take their corresponding $Y_{\rm T}=\{0.06,\; 0.02,\; 0.07,\; 0.93\}$ and Fermi momentum $p_{\rm F}= \{146, 50, 160, 373\}$ in $\rm MeV$ as calculated in~\cite{Bell:2019pyc} using the unified equation of state (EoS) BSk24 of the Brussels-Montreal model~\cite{Pearson:2018tkr}. We take $M_\star=1.5~{\rm M_\odot}$ and $R_\star=12.6~{\rm km}$ to be consistent with the calculation of $Y$ and $p_\text{F}$ in~\cite{Bell:2019pyc}. 

As an approximation, we take the volume-averaged values for $\langle n_\text{T} \rangle$, $Y_\text{T}$ and $p_\text{F}$ over the core. We have estimated the maximum deviation in our projected cut-off bounds possible due to radial variations of those quantities and different choices of EoS~\cite{Bell:2019pyc,Pearson:2018tkr, Garani:2018kkd}. These deviations may at most lead to an ${\cal O}(1)$ change in our projected sensitivities for neutrons and electrons. Detailed discussion of these variations is deferred to the section on uncertainties in the end. As we will also show later, projected bounds due to electrons in neutron star could be several orders of magnitude stronger than DM direct detection limits, thus a small $\mathcal{O}(1)$ change does not affect our main results.

We recover the usual form of ${f}$ from Eq.~\eqref{eqn:fullf} for nonrelativistic targets. As $p_{\rm F}\rightarrow 0$, the differential cross section becomes independent of $p_\mu$ and $v_\text{M\o l}\rightarrow v_\text{esc}$, also the Pauli blocking step function $\rightarrow 1$. These imply $\int|\vec{p}|^2\, d|\vec{p}|\,d\Omega_{\rm NS}\,/\,V_{\rm F}\rightarrow 1$. Assuming that a single scatter deposits at least $E_{\rm halo}$, Eq.~\eqref{eqn:fullf} gives $f=\int d\Omega_\text{CM}\left({d\sigma}/{d\Omega}\right)_\text{CM}/(\left<n_{\rm T}\right>v_\text{esc}\Delta t)^{-1}$, a well-known result, where the denominator is the geometric cross section.

%%%%%%%%%
\begin{figure*}[t!]
	\centering
	\includegraphics[width=0.45\textwidth]{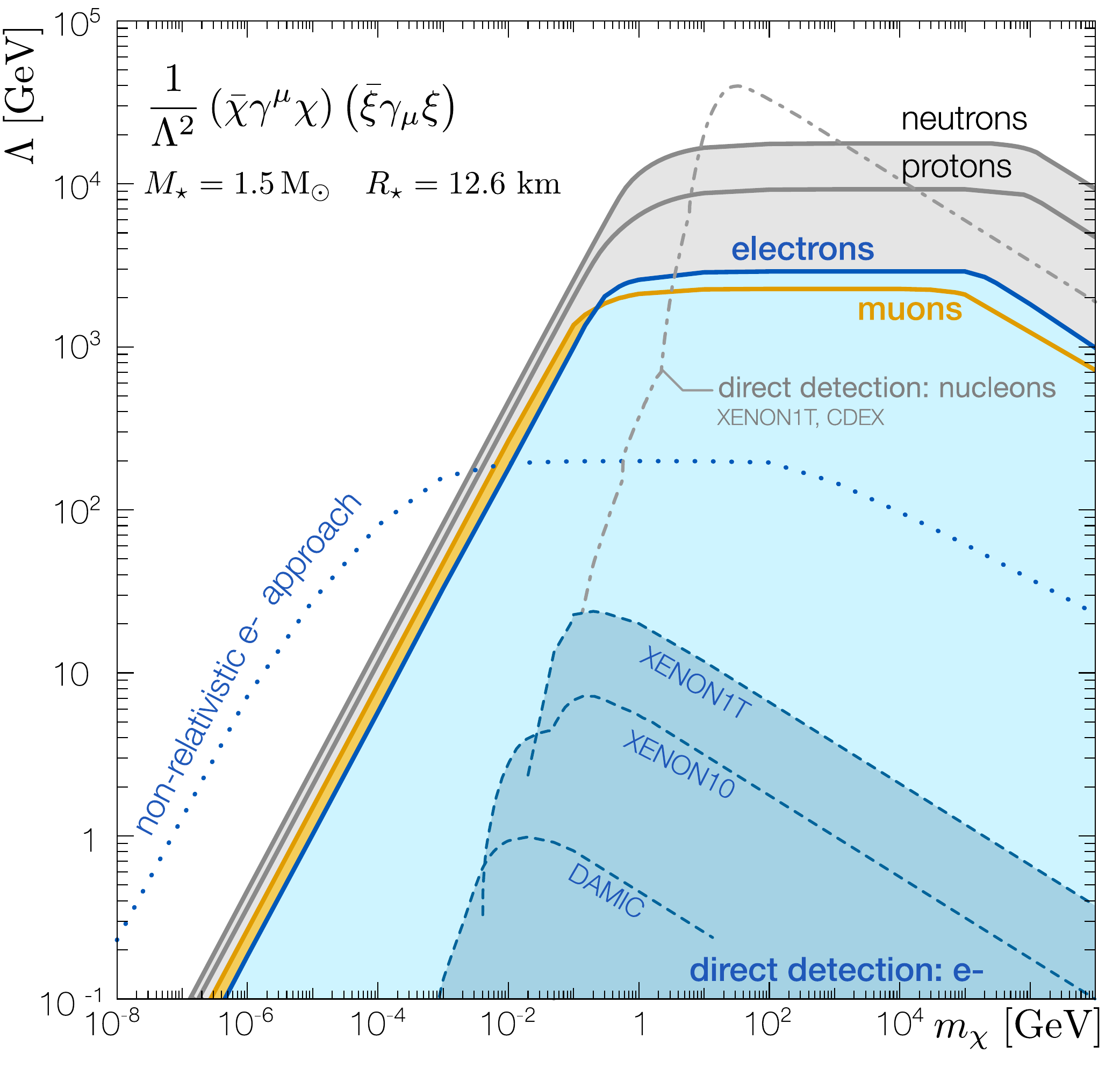} \quad 
	\includegraphics[width=0.45\textwidth]{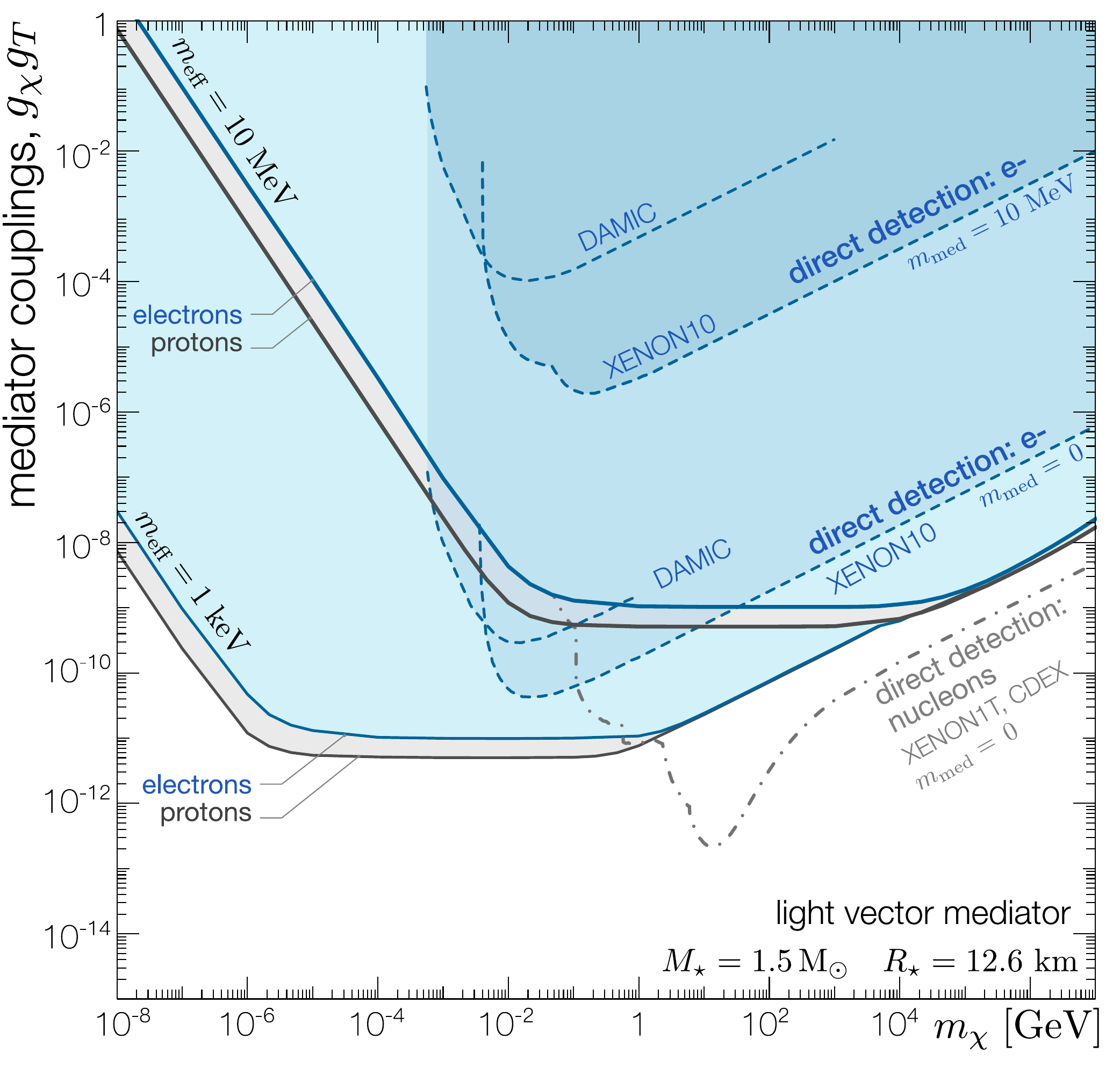}
	\caption{%
	Projected sensitivities from \NS heating for vectorial interactions of Dirac DM with Standard Model fermions (solid), together with Earth-based direct detection constraints (dashed)~\cite{Angle:2011th,Essig:2017kqs,Aguilar-Arevalo:2019wdi,Emken:2019tni,Aprile:2018dbl,Aprile:2019xxb,Aprile:2019jmx,Liu:2019kzq}. 
	{\em Left}: A heavy mediator scenario characterized by a cutoff scale $\Lambda$ for capture by various \NS constituents. The dotted line shows a non-relativistic calculation that underestimates (overestimates) the sensitivity above (below) the electron Fermi energy (electron mass).
    {\em Right}: A light mediator scenario for the capture by electrons and protons, with sensitivities displayed for the product of the mediator's couplings to DM and Standard Model fermions. The direct detection constraints here assume the mediator mass to be massless or $10$~MeV; the massive mediator lines are recast using the experimental bounds corresponding to the form factor $F_{\rm DM}=1$~\cite{Angle:2011th,Essig:2017kqs,Emken:2019tni,Aguilar-Arevalo:2019wdi}.
    In both panels, the colored regions correspond to $f=1$ ($T_\star=1600~{\rm K}$) as estimated from Eq.~\eqref{eqn:fullf}. Projected sensitivities are stronger if we take lower $f$, i.e. lower $T_\star$, corresponding to longer observation times.
 }

\label{fig:main}
\end{figure*}
%%%%%%%%%

%%%%%%%%%

\vspace{.5em}
{\noindent\bf DM model with a heavy mediator.} We apply our framework to estimate sensitivities from \NS heating for representative DM models and compare them with limits from DM direct detection experiments. We assume the DM candidate is a Dirac fermion ($\chi$) that couples to Standard Model fermions ($\xi$) through an effective vectorial operator, $(\bar{\chi}\gamma_\mu\chi)(\bar{\xi}\gamma^\mu \xi)/\Lambda^2$. 
 We explore spin-0 DM and other interactions structures in a companion paper~\cite{Joglekar:2020liw}.
The sensitivity of the proposed search is an upper limit on $\Lambda$.

Fig.~\ref{fig:main} (left) shows our projected sensitivities to the cutoff scale $\Lambda$ vs the DM mass $m_\chi$, obtained numerically, for the target fermions $\xi = e^-,\mu^-, p^+$ and $n$. The upper boundaries correspond to ${f}$ = 1, or signal temperature $T_\star=1600$~K. Stronger sensitivities could be obtained for ${f} < 1$, corresponding to smaller $T_\star$.

The plot demonstrates three distinct regimes: 
($i$) For $m_\chi\gtrsim1~{\rm PeV}$, the sensitivities decrease as the DM mass increases further. In this region, DM becomes so massive that multiple scatterings ($N_{\rm hit}>1$) are required for successful capture, suppressing the capture probability, as indicated in Eq.~\eqref{eqn:fullf}. 
($ii$)  For $p_{\rm F}\lesssim m_\chi\lesssim1~{\rm PeV}$, there are plateaus insensitive to the DM mass. In this mass range, the momentum transfer is typically larger than the Fermi momentum and Pauli blocking is unimportant. In addition, the cross section is almost independent of the DM mass. Thus the projected upper limits on $\Lambda$ are nearly constant over $m_\chi$. The electron capture sensitivity to $\Lambda$  is more than a factor $10$ stronger than the one estimated with a nonrelativistic treatment~\cite{Bell:2019pyc}.
($iii$)
For light DM, $m_\chi\lesssim p_{\rm F}$, the sensitivities decrease for all targets, due to a combined effect of Pauli blocking and suppression of the cross section, as we will discuss later. In this regime, the nonrelativistic treatment of the electrons overestimates the capture probability.

For comparison, we show constraints from DM direct direction experiments based on both electron~\cite{Angle:2011th,Essig:2017kqs,Emken:2019tni,Aguilar-Arevalo:2019wdi} and nuclear recoils~\cite{Aprile:2018dbl,Aprile:2019jmx,Aprile:2019xxb,Liu:2019kzq}. Remarkably, for light DM with $m_\chi\sim10~{\rm MeV}\textup{--}10~{\rm GeV}$, the \NS bound on $\Lambda$ can be a factor of $100$ stronger than electron recoil limits. Furthermore, \NS heating may probe a broader DM mass range not covered by direct detection for electron recoils, as well as nuclear recoils~\cite{Raj:2017wrv}. If DM couples to both electrons and nucleons equally, the limit on $\Lambda$ will be mainly set by DM--neutron/proton scatterings. On the other hand, for leptophilic DM, capture by electrons is the strongest mode of neutron star heating for $m_\chi>p_\text{F}$. We have checked that this is true even after taking into account loop-induced interactions of leptophilic DM with nucleons, in contrast to earlier results using the nonrelativistic approach~\cite{Bell:2019pyc}.

To further understand the scaling features in Fig.~\ref{fig:main} (left), we explore the scattering kinematics in more detail. The scattering cross section scales as $(d\sigma/d\Omega)_{\rm CM} \propto m^2_\chi E^2_p/(s\Lambda^4)$, where $E_p$ is the target energy in the \NS frame and $s$ is the Mandelstam variable. In the nonrelativistic limit, $E_p\approx m_{\rm T}$ and $s\approx(m_\chi+m_{\rm T})^2$, and $(d\sigma/d\Omega)_{\rm CM}$ reduces to the well-known form $(m_\chi m_{\rm T})^2/(m_\chi+m_{\rm T})^2\Lambda^4$. The DM energy and momentum in the \NS frame are $E_k=\gamma_{\rm esc}\ m_\chi$ and $|\vec{k}|=v_{\rm esc}\gamma_{\rm esc} m_\chi$ respectively, where $\gamma_{\rm esc}=1.24$ and $v_{\rm esc}=0.6$. For the electrons, these are $E_p\approx p_F$ and $|\vec{p}|=p_F$ respectively, as the electron Fermi momentum $146~{\rm MeV}$ is much larger than its mass $0.51~{\rm MeV}$, i.e., electrons in the \NS core are ultrarelativistic. 

Consider the heavy DM mass region, where $m_\chi\gg p_{\rm F}$ and Pauli blocking is unimportant. For the DM--electron system, $s=(E_k+E_p)^2-(\vec{k}+\vec{p})^2\approx E^2_k-k^2=m^2_\chi$. Thus, the scattering cross section scales as $(d\sigma/d\Omega)_{\rm CM}\propto p^2_{\rm F}/\Lambda^4$. Compared to the nonrelativistic approach in the \NS frame, where $(d\sigma/d\Omega)_{\rm CM}\propto m^2_e/\Lambda^4$, the actual cross section is a factor of $(p_{\rm F}/m_e)^2\sim10^5$ larger. Thus, the actual \NS sensitivity on $\Lambda$ is more than one order of magnitude stronger than estimated previously with nonrelativistic approach~\cite{Bell:2019pyc}, as indicated in Fig.~\ref{fig:main} (left). For the other targets, $(p_{\rm F}/m_{\rm T})^2<1$, the nonrelativistic approximation is valid. 

For light DM $m_\chi\ll p_{\rm F}$, the reach shown in Fig.~\ref{fig:main} (left) scales as $\Lambda\propto m^{3/4}_\chi$ for all targets, which can be understood as follows. For the nonrelativistic targets $n$, $p^+$ and $\mu^-$, DM energy loss has a weak dependence on the scattering angle, and the Pauli blocking factor scales as $m_\chi$. 
Moreover, the cross section $\propto m^2_\chi/\Lambda^4$. Thus, the capture probability ${f}\propto m^3_\chi/\Lambda^4$. For the ultrarelativistic electrons, $s=(E_k+E_p)^2-(\vec{k}+\vec{p})^2\approx2(E_kE_p-\vec{k}\cdot\vec{p})\propto m_\chi p_{\rm F}$, resulting in $(d\sigma/d\Omega)_{\rm CM}\propto m_\chi p_{\rm F}/\Lambda^4$. Since the energy loss only occurs for CM frame forward scatterings in this case, there is an additional suppression in the phase space $\propto m_\chi$, which is not present for the nonrelativistic targets. Thus the Pauli blocking factor scales as $m^2_\chi$, and we again have ${f}\propto m^3_\chi/\Lambda^4$ for the electron target. Note the nonrelativistic approach for electrons overestimates the sensitivity for $m_\chi< m_e$, because it does not take into account the fact that it is much harder to transfer energy to an ultrarelativistic electron than one at rest.

\vspace{.5em}

{\noindent\bf DM model with a light mediator.} We consider a DM model with a light vector mediator and corresponding scattering cross section
\begin{align}
\left(\frac{d\sigma}{d\Omega}\right)_{\rm CM}&\propto \frac{g^2_\chi g^2_{\rm T} m_\chi^2 E_p^2}{s(m^2_{\rm eff}+|\vec{q}|^2_{\rm CM})^2}~,
\label{eqn:diffxsectlight}
\end{align}
where $g_\chi$, $g_{\rm T}$ are the mediator's couplings to DM and the target respectively, $m_{\rm eff}$ is the in-medium effective mediator mass. This effective mass is simply the mediator mass $m_\text{eff} = m_\text{med}$, when $m_\text{med}$ is larger than the inverse of the Debye length that we estimate as~\cite{McDermott:2010pa,An:2013yfc,Redondo:2013lna,Chang:2016ntp,Hardy:2016kme,Knapen:2017xzo,Lin:2019uvt}
\begin{align}
\lambda_{\rm D}^{-1}\sim e\sqrt{\frac{n_e}{T_{\rm eff}}}\sim e\sqrt{\frac{n_e}{p_{\rm F}}}\approx 10~\text{MeV},
\end{align}
where $T_{\rm eff}$ is the effective temperature for Thomas–Fermi screening; $T_{\rm eff}\approx p_{\rm F}$. In deriving potential constraints from neutron stars, we take $m_{\rm eff}=\lambda^{-1}_{\rm D}\approx10~{\rm MeV}$. For reference, we also show the reach for $m_{\rm eff} = 1~{\rm keV}$.

Fig.~\ref{fig:main} (right) shows our sensitivities to $g_\chi g_{\rm T}$ for electron and proton targets. We compare to direct detection limits for DM with massless and 10~MeV mediators. For mediators in this range, we estimate that the neutron star heating reach is represented by the $m_{\rm eff}=10~\text{MeV}$ curves. For leptophilic mediators with a mass of $10$~MeV, the neutron star reach for $g_\chi g_T$ is orders of magnitudes stronger with respect to current bounds from terrestrial direct detection probes for the entire range of accessible DM masses. For the limit of massless mediators, the \NS kinetic heating reach for $m_{\rm eff}=10~{\rm MeV}$ is stronger than Earth-based detectors for DM masses lighter than $1~{\rm MeV}$ and heavier than $100~{\rm GeV}$. If DM couples to $e^-$ and $p^+$ equally, the combined bounds on $g_\chi g_{\rm T}$ would be at most stronger by a factor of $\sqrt{2}$.

As shown in Fig.~\ref{fig:main} (right), the projected reach changes slope when $m_\chi\approx m_{\rm eff}$ for both electron and proton targets. For $m_\chi\ll m_{\rm eff}\ll p_{\rm F}$, as seen from Eq. 4, ${f}\propto g^2_\chi g^2_{\rm T}m^3_\chi/m^4_{\rm eff}$ and the reach on $g_\chi g_{\rm T}\propto m^{-3/2}_\chi$. This is similar to the heavy-mediator model in the region of $m_\chi\ll p_{\rm F}$. While for $m_\chi>m_{\rm eff}$, one finds plateaus where the reach is constant with respect to $m_\chi$, and they extend towards a lower DM mass range, compared to the heavy-mediator model. As $m_\chi$ drops below $p_{\rm F}$, Pauli blocking suppresses the scattering phase space and reduces the capture probability. However, for the light-mediator model, the scattering cross section is enhanced by a small momentum transfer. These two competing effects reach a balance, resulting in the plateaus shown in Fig.~\ref{fig:main} (right).

To see this, observe that in Eq.~\eqref{eqn:diffxsectlight}, the momentum transfer $|\vec{q}|_\text{CM}$ below $m_{\rm eff}$ can not significantly enhance the differential cross section. Consider the expression for $|\vec{q}|^2_\text{CM}=2 |\bar{k}|^2_{\rm CM}(1-\cos\psi)$, where $\psi$ is the scattering angle in the CM frame. Let $|\vec{q}|_\text{CM}\sim m_{\rm eff}$ for $\psi=\psi_0$. Neglecting sub-dominant contributions to Eq.~\eqref{eqn:fullf} from the region $\psi>\psi_0$, the phase-space integral is $\int^1_{\cos\psi_0} d\cos\psi'\sim m_{\rm eff}^2/|\vec{k}|^2_{\rm CM}$. The allowed phase space is also suppressed in the magnitude of $|\vec{p}|$ as $|\vec{q}|_{\rm CM}/p_{\rm F}\sim m_{\rm eff}/p_{\rm F}$. Putting these factors together with Eq.~\eqref{eqn:diffxsectlight}, we have
\begin{align}
{f}
\propto 
\frac{g^2_\chi g^2_{\rm T}m_\chi^2 E^2_p}{sm_{\rm eff}^4}
\cdot
\frac{m_{\rm eff}^2}{|\vec{k}|^2_{\rm CM}}
\cdot
\frac{m_{\rm eff}}{p_{\rm F}}
\propto
\frac{g^2_\chi g^2_{\rm T}}{p_{{\rm F}}m_{\rm eff}},\label{eqn:flight}
\end{align}
where we use $s|\bar{k}|^2_\text{{CM}}\propto m_\chi^2 E_p^2$. Thus, ${f}$ is not sensitive to $m_\chi$ in this region. As we increase $m_\chi$, the cross section is suppressed by a high momentum transfer, and multiple scatterings become relevant; both effects lead to a small capture probability, resulting in weak reaches.

We note that for $m_\chi > m_{\rm eff}$ it is possible for incident DM to emit a mediator via bremmstrahlung and slow down, however the rate for this is expected to be negligible given the small $g_\chi$ couplings to which we are sensitive (Fig.~\ref{fig:main}) and the phase space suppression with respect to the scattering cross section.

\vspace{.5em}
{\noindent\bf Uncertainties.} In this section, we estimate the maximum deviation possible in our results for the projected reach of $\Lambda$, due to the radial variation of baryon density, BSk functional, $M_\star$, and $R_\star$. The exact calculation of these effects is beyond the scope of this paper and is deferred to future work. 
We also note that the BSk functionals from~\cite{Pearson:2018tkr} used in this paper only take into account the four target species considered above as neutron star constituents, and neglect the possible presence of any exotic phases of matter.

From Eq.~\eqref{eqn:fullf}, we observe that possible sources of uncertainties in our projected sensitivities are the baryon density, the $p_\text{F}$-dependence of the phase space integral, $Y_\text{T}$ and $\Delta t$. 
Given an equation of state functional, and a ($M_\star$, $R_\star$) pair predicted by it, the baryon density sets the values of $Y_\text{T}$ and $p_\text{F}$.
The baryon density itself varies in the core as a function of distance from the center. However, by significantly varying $M_\star$ and $R_\star$, a wide range of average baryon densities for the core can be obtained. This range is greater than the deviation from average baryon density within the core for a fixed configuration. This is because typically, the baryon number density remains relatively constant for at least half to two thirds of the radius.

%Changing to a different $M_\star$ and $R_\star$ configuration can lead to a significant change in the average baryon density of the core. By significantly varying $M_\star$ and $R_\star$, this change can be made larger than the deviations from the core average density for a fixed configuration of $M_\star$, $R_\star$, and a functional. This is because typically, the baryon number density remains relatively constant for at least half to two thirds of the radius. 
%Therefore, for a fixed configuration of a functional, $M_\star$, and $R_\star$, the deviation of baryon density from its average over the core is relatively small compared to the change in average baryon density induced by significant change of configuration. 

Hence, to estimate the maximum variation in our results, we consider two extreme average densities, allowed amongst all the valid $M_\star$ and $R_\star$ configurations of BSk22, BSk24, BSk25 and BSk26 functionals. Consequently, for high mass ($2.16\,M_\odot$) -- small radius ($11$~km) configuration, the average core baryon density is about $0.61$~fm$^{-3}$ and for low mass ($0.3\,M_\odot$) -- large radius ($13$~km) configuration, it is about $0.05$~fm$^{-3}$~\cite{Pearson:2018tkr,Bell:2019pyc}. For the dense configurations, the central baryon number density can go as high as $0.95$~fm$^{-3}$. Therefore, we consider the range $0.05$~fm$^{-3}$ to $0.95$~fm$^{-3}$ of the baryon number density for our uncertainty estimation. 
%Thus, the density varies by a factor of $4$ to $5$ compared to the density used in our results in Fig~\ref{fig:main}.
The corresponding ranges of values for $Y_\text{T}$ and $p_\text{F}$ for each target species can be obtained from~\cite{Pearson:2018tkr}. Thus, we find that the baryon density, $Y_\text{T}$, and $p_\text{F}$ vary by a factor of $< 5$ with respect to the ones considered in our results. 

Substituting all these quantities in Eq.~\eqref{eqn:fullf} and taking $1/4^\text{th}$ power, we estimate the width of uncertainty bands for the projected sensitivities on our EFT cutoff. 
For all species, the upper end of the band is a factor of $1.8$ times the values in the left panel of  Fig.~\ref{fig:main}. The only exception is sensitivity to electron scattering in the heavy DM region, where the band extends up to 3 times the reach in $\Lambda$ shown. 
The lower end of the band differs according to the target species. 
For neutrons it is at most a factor $1.7$ lower than the values in Fig.~\ref{fig:main}, while for electrons it could be a factor of at most $2.5$. 
If the central density of the neutron star configuration falls below that needed for having non-zero muon abundance, then the DM capture via muons is not possible. 
For configurations with sufficiently low densities, i.e., core average density below $0.12$~fm$^{-3}$, the muon bounds are significantly weakened.

The neutron and electron bands are well separated in the heavy DM region, but overlap in the light DM region. For configurations with densities higher than those used in Fig.~\ref{fig:main}, the electron bound in heavy DM region will move closer to the neutron bound. This is because higher Fermi momentum helps heavy DM capture by electrons unlike in the case of nucleon targets. For sufficiently high densities, electrons maintain their dominance over muons in the heavy DM region for the same reason. For light DM, the bands for electrons and neutrons overlap, with neutrons generally exhibiting slightly stronger bound compared to electrons for any given configuration. Muon targets provide higher sensitivity compared to electrons in the light DM region as seen in the left panel of Fig.~\ref{fig:main}. For configurations with sufficiently high baryon density, the electron and muon bounds remain comparable. However, for configurations with low baryon density, where the abundance of electrons in the central region strongly dominates over that of muons, the tables are turned and electrons start dominating in the light DM region as well.

We have assumed $\Delta t=2R_\star$. The number of DM particles following the paths with $\Delta t>2R_\star$ are an $\mathcal{O}(1)$ fraction of the total flux through the star. The resultant underestimation of the capture efficiency is of course mitigated by the overestimation from shorter paths with $\Delta t<2R_\star$ by a small $\mathcal{O}(1)$ factor.
Some target species like protons or muons only capture the DM up to a certain radial distance inside the core for certain configurations of $M_\star$ and $R_\star$, shrinking $\Delta t$ by a small $\mathcal{O}(1)$ factor. 
In the end, the uncertainty resulting from these factors in the sensitivity to $\Lambda$ is suppressed since $\Lambda\propto f^{1/4}$. 
We find that the uncertainty in our reach in $\Lambda$ due to the variation in $\Delta t$ is at most $\mathcal{O}(10\%)$. 

\vspace{0.3in}

\vspace{.5em}
{\noindent\bf Conclusions.} We have studied relativistic capture of DM by electrons in a \NS and developed a formalism to calculate the capture probability. It is manifestly Lorentz invariant and incorporates relativistic scattering kinematics, Pauli blocking, and the effect of multiple DM--electron scatters during stellar transit. We further applied the formalism to explore the sensitivities to parameter space of two benchmark DM scenarios and compared them with direct detection limits. The Lorentz-invariant capture probability can be five orders of magnitude larger than the traditional non-relativistic approach. This makes \NS heating one of the most promising testing grounds for probing leptophilic DM models. In the future, we could apply our formalism to other DM models~\cite{Raj:2017wrv,Tulin:2017ara,Alvarez:2019nwt,McKeen:2020vpf} and different capture scenarios~\cite{Reddy:2002ri,Bertoni:2013bsa,Acevedo:2019agu}. It is also interesting to investigate the discovery potential of old neutron stars using upcoming radio telescopes and infrared surveys, see., e.g,~\cite{FAST,CHIME,SKA,spergel2013wfirst24short}.

\vspace{0.3in}
  
\begin{acknowledgments}
We thank Nicole Bell, Joe Bramante, David Morrissey, Tongyan Lin, and Ethan Villarama for useful discussions. This work is supported by the U.~S.~Department of Energy under Grant No.~de-sc~0008541 (AJ, PT, HBY), and the Natural Sciences and Engineering Research Council of Canada (NSERC) (NR). 
TRIUMF receives federal funding via a contribution agreement with the National Research Council Canada. This work was also performed in part at the Aspen Center for Physics (NR, PT), which is supported by National Science Foundation grant PHY-1607611. A part of this work was also completed at Kavli Institute for Theoretical Physics (AJ, HBY), which is supported in part by the National Science Foundation under Grant No. NSF PHY-1748958.
\end{acknowledgments}

\bibliography{refs}

\end{document}